\documentstyle[11pt,newpasp,twoside,epsf]{article}
\def\edcomment#1{\iffalse\marginpar{\raggedright\sl#1\/}\else\relax\fi}
\reversemarginpar

\def\be{\begin{equation}}
\def\ee{\end{equation}}
\def\ba{\begin{eqnarray}}
\def\ea{\end{eqnarray}}
\def\go{\mathrel{\raise.3ex\hbox{$>$}\mkern-14mu
             \lower0.6ex\hbox{$\sim$}}}
\def\lo{\mathrel{\raise.3ex\hbox{$<$}\mkern-14mu
             \lower0.6ex\hbox{$\sim$}}}

\def\bE{{\bf E}}
\def\bB{{\bf B}}

\def\hatk{{\hat {\bf k}}}

\begin{document}
\title{Matter and Radiation in Superstrong Magnetic Fields and
Thermal Emission from Neutron Stars}
 \author{Dong Lai and Wynn C.G. Ho}
\affil{Department of Astronomy, Cornell University, Ithaca, NY 14853, USA}

\begin{abstract}
Thermal surface emissions have now been detected from more than a
dozen isolated neutron stars, including radio pulsars, radio-quiet neutron 
stars and magnetars. These detections can potentially 
provide important information on the interior physics, magnetic fields, 
and surface composition of neutron stars. Understanding the properties of
matter and radiative transfer in strong magnetic fields is essential for the
proper interpretation of the observations. We review current theoretical
works on modeling magnetized neutron star atmospheres/surface
layers, discussing some of the novel properties of matter and 
radiative transfer in strong magnetic fields. Of particular interest 
is the effect of the strong-field vacuum polarization, which dramatically
changes the radiative transfer and the emergent X-ray spectra 
from magnetars.
\end{abstract}

\section{Introduction}

Neutron stars (NSs) are born in the core collapse and subsequent
supernova explosion of massive stars and begin their lives at
high temperatures, $T\ga 10^{11}$~K. As they cool over the next
$10^5$--$10^6$~years, they act as sources of soft X-rays
with surface temperatures $\ga 10^5$~K. The cooling history of the NS depends
to a large extent on poorly constrained interior physics such as the nuclear
equation of state, superfluidity, and magnetic field (see, e.g.
Prakash et al.~2001; Yakovlev et a.~2001 for review). 
Already in the 1960s, it was recognized that one way to study NSs is through
observing their surface emissions in X-rays. This of course is very difficult
because of the small NS size. Thus detecting NSs' surface radiation
has been a long-term goal of high energy astrophysics for more than 30 years.
Thanks to the advance in X-ray telescopes (starting from {\it Einstein},
but most notably {\it ROSAT} and the more recent {\it ASCA}, 
{\it Chandra} and {\it XMM-Newton}),
such detection has become a reality. By now there are secured detections of
surface emissions from more than a dozen isolated NSs, including 
radio pulsars (e.g.,  PSR~B1055-52, B0656+14, Geminga, Vela), 
radio-quiet NSs (e.g., RX J1856.5-3754, RX J0720.4-3125),
and magnetars (AXPs and SGRs) (see Becker \& Pavlov 2002 for a review). 
These detections are important because they probe the near vicinity 
and interior of NSs: From the spectrum we can measure the surface temperature,
and thus constrain the NS radius and cooling history, which in turn depend on
the interior physics; we can also potentially measure the surface composition
and magnetic fields of NSs\footnote{For example, the absorption features
(around 0.7~keV and 1.4~keV) detected in the radio-quiet NS 1E~1207-5209 by
{\it Chandra} and {\it XMM-Newton} (Sanwal et al.~2002; Mereghetti et al.~2002;
Hailey \& Mori 2002) have been interpreted as due to atomic level transitions
of heavy elements in a $10^{11}$--$10^{12}$~G field, or due to He$^+$
transitions in a $10^{14}$~G field. However, the true nature of these lines is
currently unclear, largely due to lack of theoretical modeling effort.}.
The thermal emission is mediated by the outmost layer of the NS.  
In order to properly interpret the current and future observations,
it is crucial to have a detailed understanding of the physical
properties of the matter and radiation in strong 
magnetic fields ($B\sim 10^{11}$--$10^{16}$~G) and to calculate
the emergent thermal radiation spectra from the NSs.

Some basics about NS atmospheres are in order. Because of the strong gravity,
the NS atmosphere is highly compressed, with scale height $0.1$--$10$~cm 
and density $\sim 0.1$--$10^3$~g/cm$^3$. Thus we are dealing with
a highly nonideal gas and effects like pressure ionization are important.
The physical properties of the atmosphere, such as the
chemical composition, equation of state, and especially the radiative
opacities, directly determine the characteristics of the thermal emission.  
While the surface composition of the NS is unknown,
a great simplification arises due to the efficient gravitational
separation of light and heavy elements.
A pure H atmosphere is expected even if a small
amount of fallback/accretion occurs after NS formation.
A He atmosphere results if H is completely burnt up,
and a heavy-element (e.g., C, O, or Fe) 
atmosphere may be possible if no fallback/accretion occurs.
The strong magnetic field makes the atmospheric plasma anisotropic and
birefringent. If the surface temperature is not too high, atoms and molecules 
may form in the atmosphere. Moreover, if the magnetic field is sufficiently
strong, the NS envelope may transform into a condensed phase with very little 
gas above it. A superstrong magnetic field will also make some QED effects
(e.g., vacuum polarization) important in calculating the surface radiation
spectrum.

\section{Matter in Strong Magnetic Fields: Brief Overview}

We now briefly review the basic properties of matter in strong magnetic
fields (see Lai 2001 and references therein). 

An electron in a uniform magnetic field $B$ gyrates around 
the field line at the frequency $\omega_{ce}=eB/(m_ec)$.
In quantum mechanics, this transverse motion is quantized into Landau levels,
with the cyclotron energy (the Landau level spacing) given by 
$\hbar\omega_{ce}=11.58\,B_{12}$~keV, and the cyclotron radius 
(the characteristic size of the wave packet) becomes
$\hat R= \left({\hbar c/eB}\right)^{1/2}=2.57\times
10^{-10}B_{12}^{-1/2}$~cm, where $B_{12}=B/(10^{12}\,{\rm G})$.
When studying matter in magnetic fields, the natural (atomic) unit
for the field strength, $B_0$, is set by 
$\hbar\omega_{ce}=e^2/a_0$, 
where $a_0$ is the Bohr radius. Thus it is convenient to 
define a dimensionless magnetic field strength $b$ via 
\be
b\equiv {B/B_0};\qquad B_0={m_e^2e^3c/\hbar^3}=2.3505\times 10^9\,
{\rm G}.
\label{eqb0}\ee
For $b\gg 1$, the electron cyclotron energy 
$\hbar\omega_{ce}$ is much larger than the typical Coulomb energy,
so that the properties of atoms, molecules and condensed matter are 
qualitatively changed by the magnetic field. 
In such a strong field regime, the usual perturbative
treatment of the magnetic effects (e.g., Zeeman splitting of 
atomic energy levels) does not apply.
Instead, the Coulomb forces act as a perturbation to the magnetic
forces, and the electrons in an atom/molecule settle into the ground Landau
level. Because of the extreme confinement ($\hat R\ll a_0$) of the electrons in
the transverse direction (perpendicular to the field), the Coulomb force
becomes much more effective in binding the electrons along the magnetic field
direction. The atom attains a cylindrical structure. Moreover, it is possible
for these elongated atoms to form molecular chains by covalent bonding along
the field direction. Interactions between the linear chains can then lead to
the formation of three-dimensional condensates.

We now discuss some basic properties of different bound states in 
strong magnetic fields, using H as an example.

{\bf (i) Atom:} For $b\gg 1$, the H atom is elongated and squeezed, 
with the transverse size (perpendicular to $\bB$) $\sim \hat R=a_0/b^{1/2}
\ll a_0$ and the longitudinal size $\sim a_0/(\ln b)$. Thus the ground-state
binding energy $|E|\simeq 0.16\,(\ln b)^2$(au) (where 1~au = 
27.2~eV; the factor
$0.16$ is an approximate number). Thus $|E|=160,540$~eV at $B=10^{12},
10^{14}$G respectively. 
In the ground state, the guiding center of the electron's 
gyro-motion coincides with the proton. The excited states of the 
atom can be obtained by displacing the guiding center away from the proton;
this corresponds to $\hat R\rightarrow R_s=(2s+1)^{1/2}\hat R$ (where
$s=0,1,2,\cdots$). Thus $E_s\simeq -0.16\,\left\{\ln [b/(2s+1)]
\right\}^2$(au).

This simple picture of the H energy levels is modified when we 
consider the effect of finite proton mass: Even for a ``stationary'' H atom,
the energy $E_s$ is changed to $E_s+s\hbar\omega_{cp}$, where 
$\hbar\omega_{cp}=6.3\,B_{12}$~eV is the proton cyclotron energy. Thus the
extra energy $s\hbar\omega_{cp}$ (which can be thought of as a ``recoil''
term) becomes increasingly important with increasing $B$. Moreover,
the effect of center-of-mass motion is nontrivial: When the atom moves
perpendicular to the magnetic field, a strong electric field is induced in its
rest frame and can significantly change the atomic structure (the ``motional
Stark effect''); indeed, the mobility of the neutral atom across the magnetic
field is limited. As a result, the dependence of the atomic energy on 
the transverse momentum is complicated. This effect leads to a large shift and
broadening of the energy levels and significant modification to the 
ionization equilibrium.

{\bf (ii) Molecules and Chains:}
In a strong magnetic field, the mechanism of forming molecules is quite
different from the zero-field case. The spins of the electrons in the atoms
are aligned anti-parallel to the magnetic field, and thus two atoms in
their ground states do not bind together according to the exclusion principle. 
Instead, one H atom has to be excited to the $s=1$ state before
combining (by covalent bond) with another atom in the $s=0$ state.
Since the ``activation energy'' for exciting an electron in the 
H atom from $s$ to $(s+1)$ is small, the resulting H$_2$ molecule is stable.
Moreover, in strong magnetic fields, stable H$_3$, H$_4$ etc. can be formed
in the similar manner. The dissociation energy of the molecule is
much greater than the $B=0$ value: e.g., for H$_2$, it is 40,350~eV at
$10^{12},10^{14}$~G respectively.

{\bf (iii) Condensed Matter:} As more atoms are added to a molecule, 
the energy per atom in a H$_n$ molecule saturates, becoming independent
of $n$; this occurs at $n\go \left[b/(\ln b)^2\right]^{1/5}$ ($\sim 3-5$
for field strengths of interest).  We then have a 1D metal. 
To obtain the basic scaling relation, we can consider a uniform cylinder 
model in which a 1D ion lattice is embedded in an electron Fermi 
sea. The radius of the cylinder and the ion spacing are of order $a\sim
Z^{1/5}b^{-2/5}$ and the energy per ``atom'' is $E\sim -Z^{9/5}b^{2/5}$ 
(where we have restored the ion charge number $Z$). By placing 
parallel chains together (with spacing $\sim a$) we form a 3D 
condensed matter (e.g., in a body-centered tetragonal lattice). 
The energy per unit cell is again of order $E\sim -Z^{9/5}b^{2/5}$.
The radius of the cell is $R\sim Z^{1/5}b^{-2/5}$, corresponding to 
the zero-pressure density $\simeq 10^3AZ^{3/5}B_{12}^{6/5}$~g~cm$^{-3}$
(where $A$ is the mass number of the ion).

{\bf (iv) Phase Diagram:} Having understood the different bound
states of H, the next question is: What is the physical condition of
the H surface layer of a NS as a function of $B$ and $T$? What is the
phase diagram? Clearly, we are dealing with a highly magnetized,
dense (0.1--$10^3$~g~cm$^{-3}$) and partially ionized plasma; further 
complications arise from the strong coupling between the center-of-mass 
motion  and the internal structure of atoms/molecules. Calculations indicate
that there are two possible regimes: (1) Under ``normal'' conditions
($B\lo 10^{14}$~G and $T\go 10^6$~K), the surface layer (photosphere)
is gaseous and nondegenerate, with a mixture of p,~e,~H atoms, and H$_2$, etc.
(depending on $B,T,\rho$; (2) Under more ``extreme'' conditions 
($T\gg 10^{14}$~G and/or $T\ll 10^6$~K), there is a phase transition from the
gaseous phase to the condensed metallic phase; as 
$B$ increases, the vapor density (above the metal) decreases, and we then have
a situation in which the surface consists of condensed metallic H from which
radiation directly comes out. The precise boundary between the two regimes 
(or the precise critical temperature) is currently uncertain (Lai \& Salpeter
1997).

\section{Neutron Star Atmosphere Modeling: Current Status}

Current zero-field NS atmospheres are based on the opacity and 
equation of state data from the OPAL project for pure H, He and Fe 
compositions (Rajagopal \& Romani 1996; Zavlin et al.~1996; G\"ansicke et
al.~2002). These works showed that the radiation spectra from light-element (H
or He), low-field ($B\lo 10^8$~G) atmospheres deviate significantly from 
blackbody. So far most studies of magnetic NS atmospheres have focused
on H and moderate field strengths of $B\sim 10^{12}$--$10^{13}$~G
(see Zavlin \& Pavlov 2002 for a review).
These models take into account the transport of different photon modes 
through a mostly ionized medium. The opacities adopted in the
models include free-free transitions and electron scattering,
while bound-free opacities are either neglected or treated in a very
approximate manner and bound-bound transitions are ignored.
Thus these magnetic atmosphere models are expected to be valid only 
for relatively high temperatures ($T\go {\rm a~few}\times 10^6$~K) 
where hydrogen is almost completely ionized. 
As the magnetic field increases, we expect these models
to break down at even higher temperatures as bound atoms, molecules 
and condensate become increasingly important. 
Models of magnetic iron atmospheres (with $B\sim 10^{12}$~G)
were studied by Rajagopal et al.~(1997). Because of the complexity
in the atomic physics and radiative transport, these Fe
models are necessarily crude. Despite some shortcomings,
the H and Fe atmosphere models have played a valuable role in assessing the
observed thermal spectra of NSs
(see Pavlov et al.~2002 for a review).

Much work remains to incorporate bound species in 
magnetic NS atmosphere models. Although the atomic
structure and radiative transitions of a atom in strong
magnetic fields are well understood (e.g. Ruder et al.~1994; 
Miller \& Neuhauser 1991;
Mori \& Hailey 2002), the complication arises from 
the effect of center-of-mass motion discussed in \S2.
Thermodynamically consistent equation of state and
opacities for a magnetized H plasma have been calculated 
by Potekhin \& Chabrier (2003). Atmosphere models based on 
this equation of state shows that the most important
transitions are from $s=0$ to $s=1$ and photoionization; both features
are significantly broadened by the center-of-mass 
motion effect (see Ho et al.~2003).

The other area of recent development concerns 
NS atmospheres in the superstrong field regime
($B\go 10^{14}$~G) (Ho \& Lai 2001,~2003,
Lai \& Ho 2002,~2003, hereafter HL01,HL03,LH02, LH03;
\"Ozel 2001, Zane et al.~2001), which we discuss in \S4. 

\section{Radiative Transfer in Superstrong Magnetic Fields: Effect of
Vacuum Polarization}

\begin{figure}
\plotfiddle{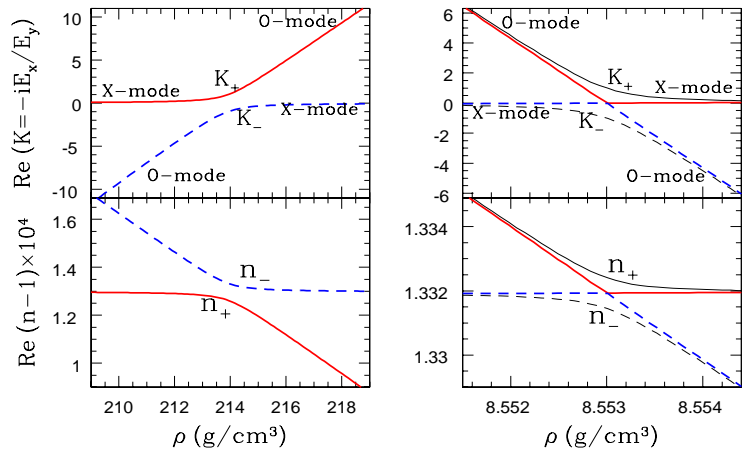}{3.0in}{0}{50}{48}{-160}{-100}
\vskip -2cm
\caption{The polarization ellipticities $K$ (upper panels) 
and refractive indices (lower panels) of the 
photon modes as functions of density near vacuum resonance for $B=5\times
10^{14}$~G, $\theta_B=45^\circ$, and $Y_e=1$. The left panels are for
$E=5$~keV, and the right panels for $E=1$~keV. On the right panels,
the light curves show the results when damping is neglected, while
the heavy lines include damping (for $T=5\times 10^6$~K). On the left 
panels, the results are indistinguishable with and without damping. 
From Lai \& Ho (2003).
}\end{figure}

Polarization of the vacuum due to virtual $e^+e^-$ pairs becomes important 
when 
\be
B\go B_Q=m_e^2c^3/e\hbar=4.414\times 10^{13}~{\rm G};
\ee
this is a genuine quantum electrodynamics effect (e.g., Adler 1971;
Heyl \& Hernquist 1997). 
Vacuum polarization modifies the dielectric property of the medium and the
polarization of photon modes, thereby altering the radiative opacities 
(e.g., Gnedin et al.~1978; Meszaros \& Ventura 1979; 
Pavlov \& Gnedin 1984;
Bulik \& Miller 1997; see Meszaros 1992 for a review). 
Of particular importance is the ``vacuum resonance'' phenomenon, which occurs
when the effects of vacuum and plasma on the polarization of the modes
``compensate'' each other. For a photon of energy $E$, the vacuum 
resonance occurs at the density 
\be
\rho_V\simeq 0.964\,Y_e^{-1}B_{14}^2E_1^2 f^{-2}~{\rm g~cm}^{-3},
\label{eq:densvp}
\ee
where $Y_e$ is the electron fraction, $E_1=E/(1~{\rm keV})$,
$B=10^{14}B_{14}$~G is the magnetic field
strength, and $f=f(B)$ is a slowly varying function of $B$ and is of order
unity (LH02 and HL03). 
For $\rho>\rho_V$ (where the plasma effect dominates the dielectric
tensor) and $\rho<\rho_V$ (where vacuum polarization dominates), the 
photon modes (for $E$ much smaller than the electron cyclotron energy
$E_{Be}$) are almost linearly polarized (see Fig.~1): the extraordinary 
mode (X-mode) has its electric field vector $\bE$
perpendicular to the $\hatk$-${\hat\bB}$ plane, while the ordinary mode 
(O-mode) is polarized along the $\hatk$-$\hat\bB$ plane
(where $\hatk$ specifies the direction of photon propagation, $\hat\bB$ 
is the unit vector along the magnetic field). Near $\rho=\rho_V$, however,
the normal modes become circularly polarized as a result of the
``cancellation'' of the plasma and vacuum effects --- both effects tend to 
make the mode linearly polarized, but in mutually orthogonal directions. 
When a photon propagates in an inhomogeneous medium, its polarization
state will evolve adiabatically if the density variation is
sufficiently gentle. Thus, a X-mode (O-mode) photon will be converted
into a O-mode (X-mode) as it traverses the vacuum resonance (see Fig.~2).
This resonant mode conversion is analogous to the MSW effect of neutrino
oscillation (e.g., Haxton 1995). For this conversion to be effective, 
the adiabatic condition must be satisfied:
\be
E\go E_{\rm ad}(B,\theta_B,H_\rho)
=1.49\,\bigl(f\,\tan\theta_B |1-u_i|\bigr)^{2/3}
\left({5\,{\rm cm}/H_\rho}\right)^{1/3}~{\rm keV},
\label{condition}\ee
where $\theta_B$ is the angle between ${\hatk}$ and ${\hat\bB}$, $u_i
=(E_{Bi}/E)^2$, $E_{Bi}$ is the ion cyclotron energy,
and $H_\rho=|dz/d\ln\rho|$ is the density scale
height (evaluated at $\rho=\rho_V$) along the ray.
For an ionized Hydrogen atmosphere, $H_\rho\simeq 2kT/(m_pg\cos\theta)
=1.65\,T_6/(g_{14}\cos\theta)$~cm, where $T=10^6\,T_6$~K is the temperature,
$g=10^{14}g_{14}$~cm~s$^{-2}$ is the gravitational acceleration, and $\theta$
is the angle between the ray and the surface normal.
We note that here we refer to the O-mode as the mode with $|K|=|E_x/E_y|\gg 1$
and the X-mode as the mode with $|K|\ll 1$ (here the $z$-axis is along $\hatk$
and the $y$-axis is in the direction of $\hat\bB\times \hatk$), thus
the name ``mode conversion''. Alternatively, one can define modes with 
definite helicity (the sign of $K=-iE_x/E_y$) such that $K$ changes
continuously as $\rho$ changes; we call these the plus-mode and 
minus-mode. Thus we may also say that in the adiabatic limit, the
photon will remain in the same plus or minus branch, but the character of
the mode is changed across the vacuum resonance. Indeed, in the 
literature on radio wave propagation in plasmas,
the nonadiabatic case, in which the photon state
jumps across the continuous curves, is referred to as ``linear mode coupling''.
It is important to note that the ``mode conversion'' effect discussed here
is not a matter of semantics. The key point is that in the adiabatic limit, 
the photon polarization ellipse changes its orientation across the vacuum
resonance, and therefore the photon opacity changes significantly
\footnote{The O-mode has a significant component of its $\bE$ field along 
$\hat\bB$ (for most directions of propagation except when $\hatk$ is 
nearly parallel to $\hat\bB$), and therefore the O-mode
opacity is close to the $B=0$ value, while the X-mode opacity 
is much smaller.}. 


\begin{figure}[h]
\plottwo{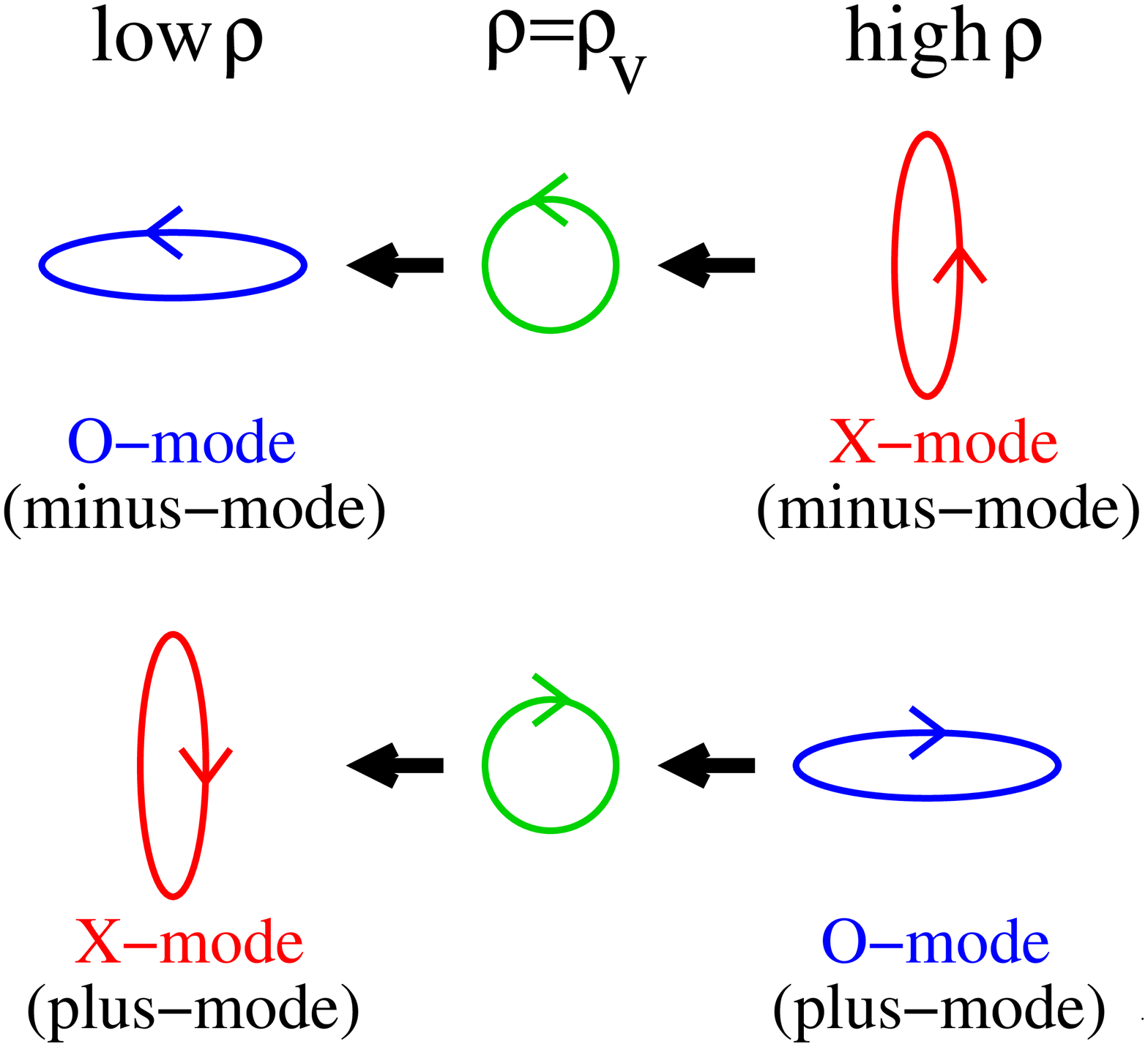}{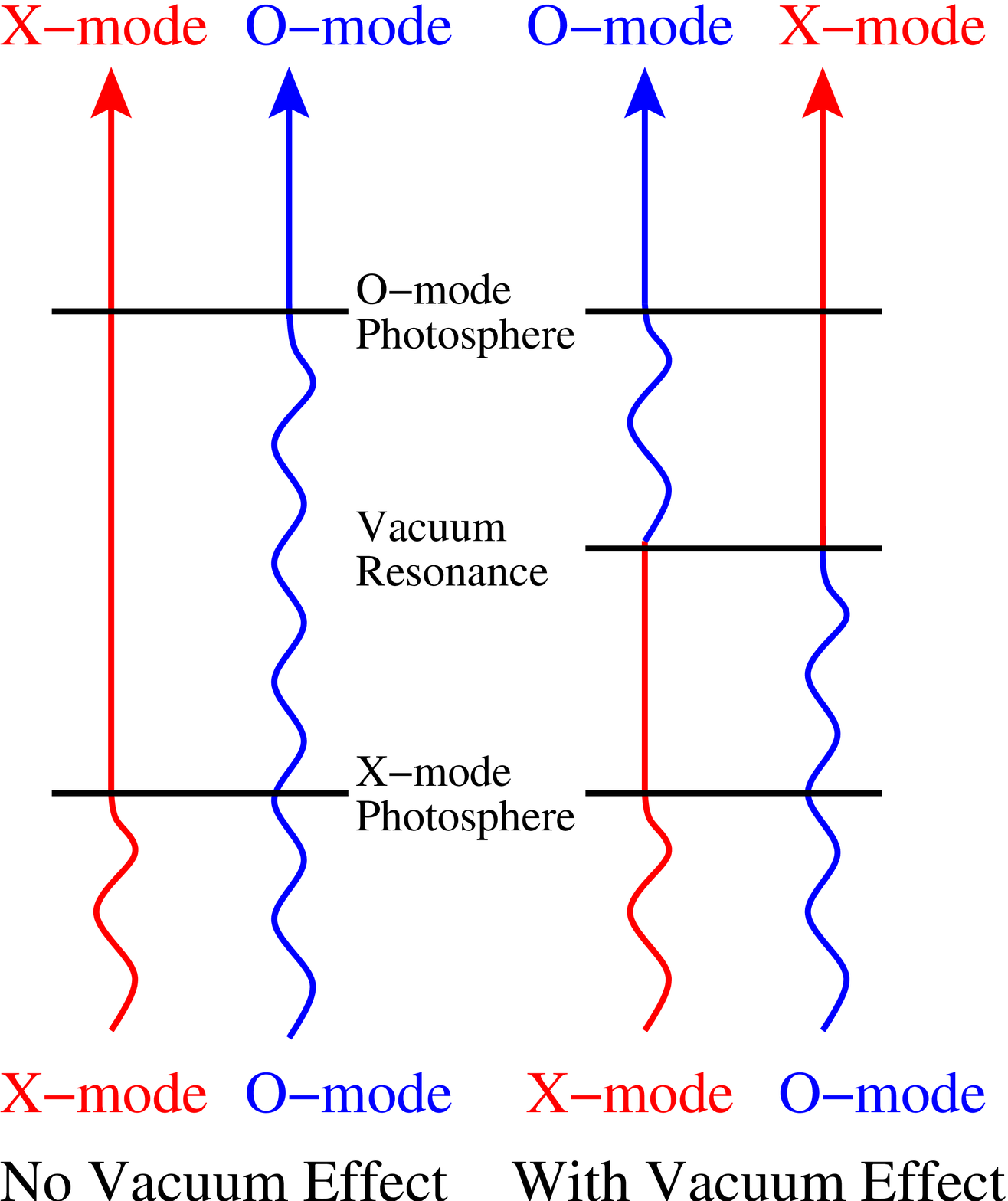}
\caption{Left panel (a schematic diagram illustrating mode conversion due to
vacuum polarization): As a minus-mode (plus-mode) photon, which manifests 
as X-mode (O-mode) at high density (for $E>E_{Bi}$), traverses the vacuum
resonance density $\rho_V$, it will stay as minus-mode (plus-mode) 
and become O-mode (X-mode) at low density if the adiabatic condition is
satisfied; in this evolution, the polarization ellipse rotates $90^\circ$
and the photon opacity changes significantly. Right panel
(a diagram illustrating how vacuum polarization-induced 
mode conversion affects radiative transfer in a magnetar atmosphere):
When the vacuum polarization effect is turned off, the X-mode 
photosphere (where optical depth $\sim 1$) lies deeper than the O-mode; 
with the vacuum polarization effect included, the X-mode effectively
decouples (emerges) from the atmosphere at the vacuum resonance,
which lies at a lower density than the (original) X-mode photosphere.
From Lai \& Ho (2003).
}\end{figure}

Because the two photon modes have vastly different opacities (see footnote 2),
the vacuum polarization-induced mode conversion can significantly affect
radiative transfer in magnetar atmospheres. The main effect of
vacuum polarization on the atmosphere spectrum can be understood
as follows. When the vacuum polarization effect is neglected, 
the decoupling densities of the O-mode and X-mode photons 
(i.e., the densities of their respective photospheres)
are approximately given by (for a H plasma) 
$\rho_O\simeq 0.42\,T_6^{-1/4}E_1^{3/2}G^{-1/2}$~g~cm$^{-3}$ and
$\rho_X\simeq 486\,T_6^{-1/4}E_1^{1/2}B_{14}G^{-1/2}$~g~cm$^{-3}$ (see LH02), 
where $G=1-e^{-E/kT}$. The vacuum resonance lies between these two photospheres
when $\rho_O<\rho_V<\rho_X$, i.e.
\be
0.66\,f\,T_6^{-1/8}E_1^{-1/4}G^{-1/4}<B_{14}<
510\,f^2\,T_6^{-1/4}E_1^{-3/2}G^{-1/2}.
\ee
When this condition is satisfied, the effective decoupling depths of
the photons are changed\footnote{For larger $B$ the vacuum resonance lies
deeper than the photospheres of both modes, while for smaller $B$
the resonance lies outside both photospheres. In both cases,
the effect of vacuum polarization on the radiative spectrum is expected
to be small, although in the latter case (smaller $B$) the polarization
of the emitted photons will be modified by the vacuum resonance.}.
Indeed, we see from Fig.~2 (right panel) that mode conversion
makes the effective decoupling density of X-mode photons (which carry the
bulk of the thermal energy) smaller, thereby depleting the high-energy
tail of the spectrum and making the spectrum closer to black-body (although
the spectrum is still harder than black-body because of nongrey opacities)
\footnote{Even when mode conversion is neglected, the X-mode
decoupling depth can still be affected by vacuum polarization.
This is because the X-mode opacity exhibits a spike feature 
near the resonance, and the optical depth across the resonance region
can be significant;  see LH02.}.
This expectation is borne out in self-consistent atmosphere modeling
presented in HL03 (see Fig.~3). 
Another important effect of vacuum polarization on the
spectrum, first noted in HL03, is the suppression of proton cyclotron lines
(and maybe other spectral lines; see Ho et al.~2003). 
The physical origin for such line
suppression is related to the depletion of continuum flux, which makes
the decoupling depths inside and outside the line similar. HL03 suggests
that the absence of lines in the observed spectra of several AXPs
(e.g., Juett et al.~2002; Tiengo et al.~2002) may be an
indication of the vacuum polarization effect at work in these systems.

\begin{figure}[h]
\plottwo{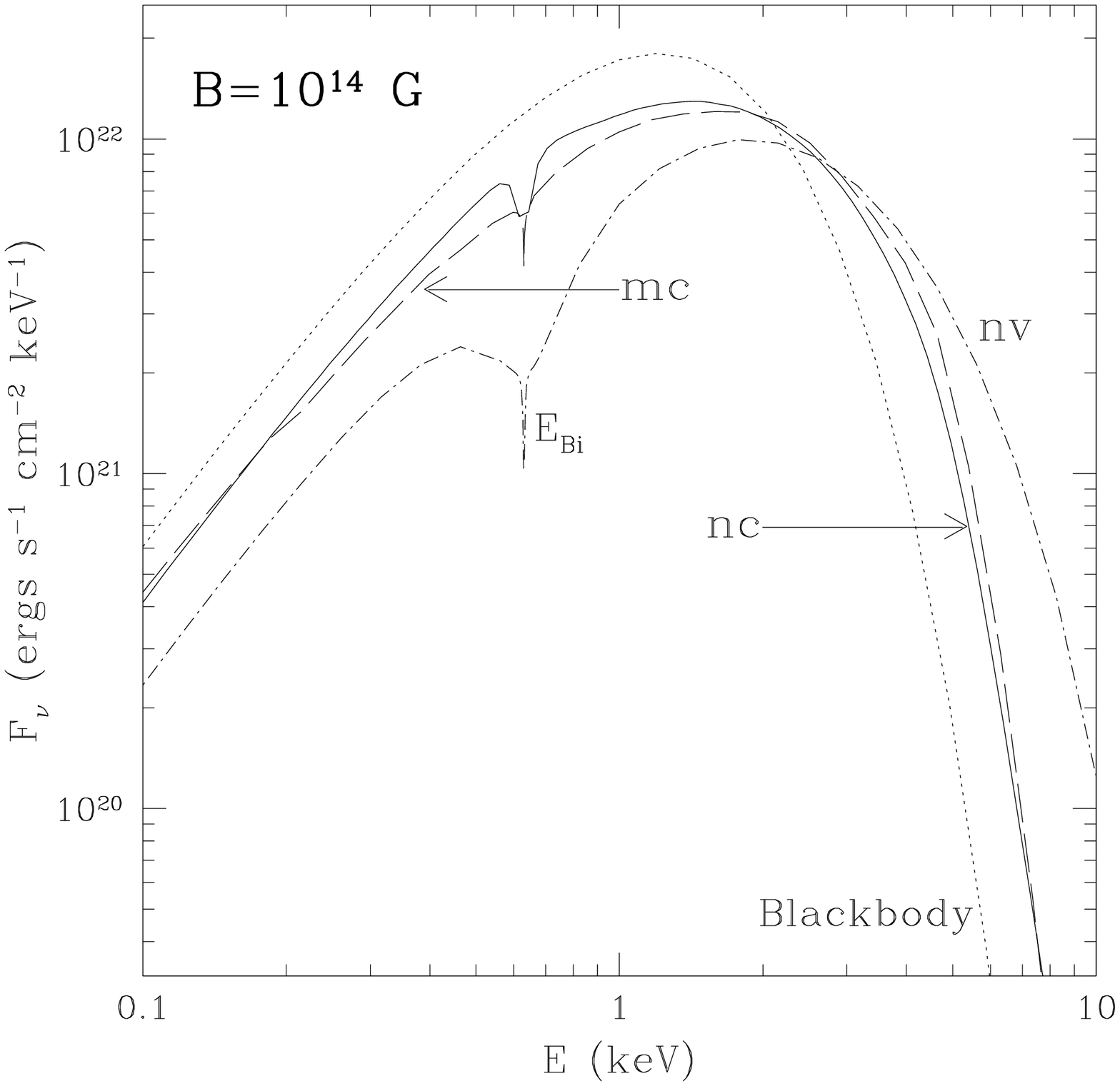}{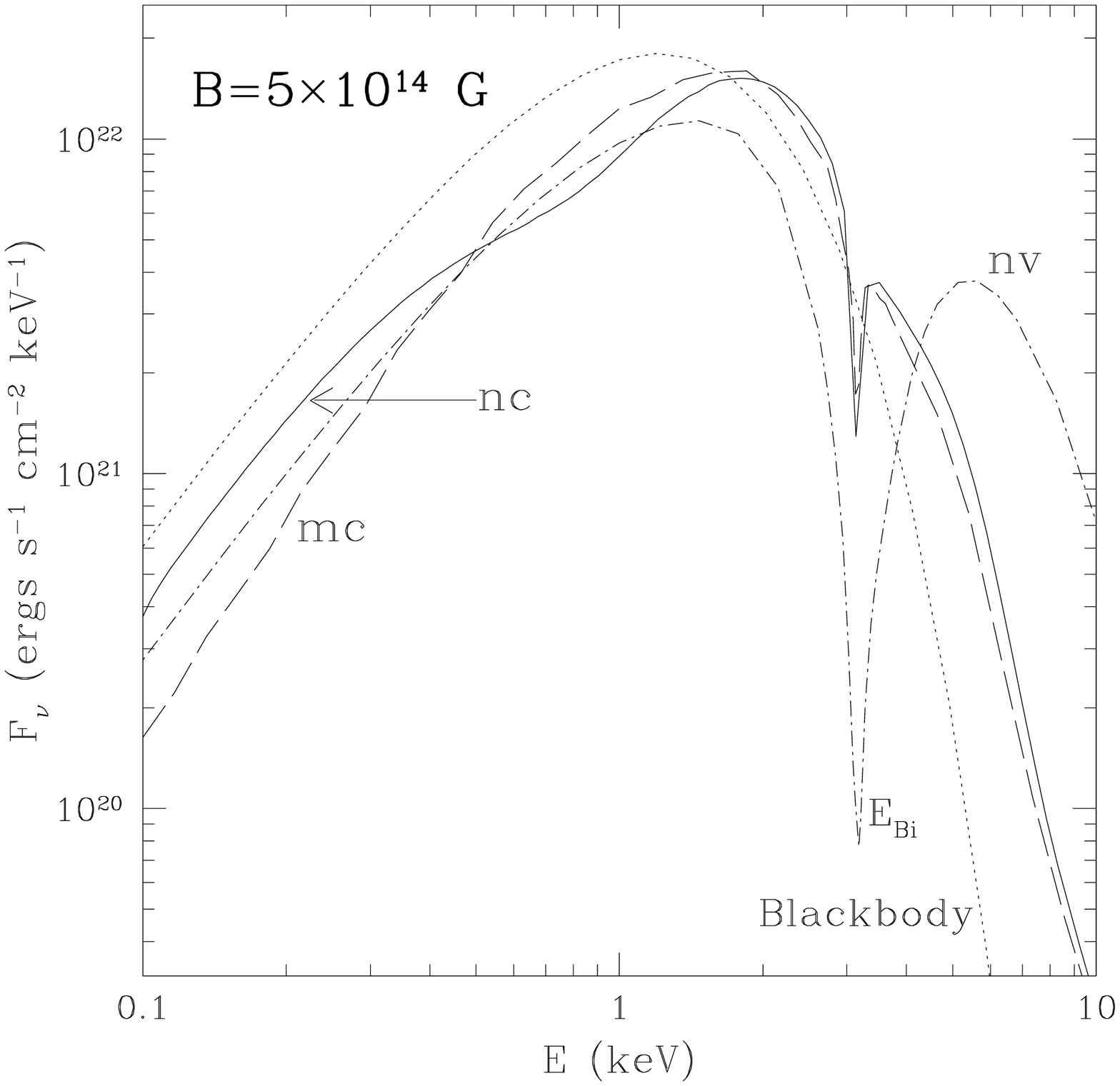}
\caption{Spectra of fully ionized H atmospheres with
$B_{14}=1$ (left) and $5$ (right), both for $T_{\rm eff}=5\times 10^6$~K,
and $\bB$ perpendicular to the surface. 
The solid line is for an atmosphere with vacuum polarization
but no mode conversion (nc),
the dashed line is for an atmosphere with complete vacuum-induced
mode conversion (mc), the dot-dashed line is for an atmosphere with no vacuum
polarization (nv), and the dotted line is the blackbody spectrum.
The $E_{Bi}=0.63\,B_{14}$~keV features are due to ion cyclotron resonance.
From HL03.
}\end{figure}

Our previous study (LH02) of the vacuum-induced mode conversion 
did not explicitly take into account of the effect of dissipation 
on the mode structure. Although this dissipative effect
is small under many situations, near the
vacuum resonance and for some directions of propagation,
the two photon modes can ``collapse'', i.e., they become identical and 
hence nonorthogonal (Soffel et al.~1983).
The analysis of LH02 breaks down near these ``mode collapse'' points.
Also, all previous studies of NS atmospheres
with strong magnetic fields rely on the modal 
description of the radiative transport. This is valid
only in the limit of large Faraday depolarization (Gnedin \& Pavlov 1974),
which is not always satisfied near the vacuum resonance, 
especially for superstrong field strengths (Lai \& Ho 2003).
More importantly, the transfer equations based on normal modes
cannot handle the cases in which partial mode coupling (conversion)
occurs across the vacuum resonance (i.e., when the adiabatic condition is
neither strongly satisfied or violated). Clearly, to properly account 
for the effects of mode collapse, breakdown
of Faraday depolarization, and mode conversion
associated with vacuum polarization, one must go beyond the 
modal description of the radiation field by formulating 
and solving the transfer equation in terms of
the photon intensity matrix (or Stokes parameters)
and including the birefringence of the plasma-vacuum medium
(see Lai \& Ho 2003). 

\acknowledgments
This work is supported in part by NASA 
grants NAG 5-8484 and NAG 5-12034, and NSF grant AST 9986740.

\end{document}